\documentclass[apsprl,longbibliography,showpacs,floatfix,superscriptaddress,twocolumn]{revtex4-2}
\usepackage[usenames,dvipsnames]{color} 
\usepackage{graphicx}
\usepackage{amsfonts}
\usepackage[figuresright]{rotating}  
\usepackage{amssymb}
\usepackage{bm}
\usepackage{amsmath}
\usepackage{mathtools}
\usepackage{psfrag}
\usepackage{multirow}
\usepackage{tabularx}
\usepackage{textcomp}
\usepackage{units}
\usepackage{lipsum}
\usepackage{xr}
\usepackage{physics}
\usepackage{soul}
\usepackage{titlesec}
\usepackage{times}
\usepackage[colorlinks=true,citecolor=blue,linkcolor=magenta,hypertexnames=false]{hyperref}
\usepackage{enumitem}

\usepackage{subfigure}

\DeclareMathAlphabet\mathbfcal{OMS}{cmsy}{b}{n}

\titlespacing*{\section}
{0pt}{1ex plus .25ex}{1ex plus 1ex}
\titlespacing*{\subsection}
{0pt}{1ex plus .25ex}{1ex plus 1ex}
\titlespacing*{\subsubsection}
{0pt}{1ex plus .25ex}{1ex plus 1ex}

\titleformat{\section}
 {\fontsize{10}{15}\bfseries }
  {\thesection}
  {1em}
  {}

\titleformat{\subsection}
  {\bfseries }
  {\thesubsection}
  {2em}
  {}

\titleformat{\subsubsection}
  {\itshape}
  {\thesubsubsection}
  {2em}
  {}

\def\beq{\begin{eqnarray}}
\def\eeq{\end{eqnarray}}

\let\baraccent=\= 
\renewcommand{\=}[1]{\stackrel{#1}{=}} 

\makeatletter

\titleclass{\subsubsubsection}{straight}[\subsection]

\newcounter{subsubsubsection}[subsubsection]
\renewcommand\thesubsubsubsection{\thesubsubsection.\arabic{subsubsubsection}}

\titleformat{\subsubsubsection}
   {\normalfont\normalsize\itshape \centering}{\thesubsubsubsection}{1em}{}
\titlespacing*{\subsubsubsection}
{0pt}{1ex plus .3ex}{1ex plus .3ex}

\makeatletter
\renewcommand\paragraph{\@startsection{paragraph}{5}{\z@}%
  {3.25ex \@plus1ex \@minus.2ex}%
  {-1em}%
  {\normalfont\normalsize}}
\renewcommand\subparagraph{\@startsection{subparagraph}{6}{\parindent}%
  {3.25ex \@plus1ex \@minus .2ex}%
  {-1em}%
  {\normalfont\normalsize}}
\def\toclevel@subsubsubsection{4}
\def\toclevel@paragraph{5}
\def\toclevel@paragraph{6}
\def\l@subsubsubsection{\@dottedtocline{4}{7em}{4em}}
\def\l@paragraph{\@dottedtocline{5}{10em}{5em}}
\def\l@subparagraph{\@dottedtocline{6}{14em}{6em}}
\makeatother

\begin{document}

\title{Topological quantum criticality from multiplicative topological phases }
\author{R. Flores-Calderón}
\affiliation{Max Planck Institute for the Physics of Complex Systems, Nöthnitzer Strasse 38, 01187 Dresden, Germany}
\affiliation{Max Planck Institute for Chemical Physics of Solids, Nöthnitzer Strasse 40, 01187 Dresden, Germany}
\author{Elio J. König}
\affiliation{Max Planck Institute for Solid State Physics, Heisenbergstra{\ss}e 1, 70569 Stuttgart, Germany}

\author{Ashley M. Cook}
\affiliation{Max Planck Institute for the Physics of Complex Systems, Nöthnitzer Strasse 38, 01187 Dresden, Germany}
\affiliation{Max Planck Institute for Chemical Physics of Solids, Nöthnitzer Strasse 40, 01187 Dresden, Germany}

\begin{abstract}
Symmetry-protected topological phases (SPTs) characterized by short-range entanglement include many states essential to understanding of topological condensed matter physics, and the 
extension to gapless SPTs provides essential understanding of their consequences. In this work, we identify a fundamental connection between gapless SPTs and recently-introduced multiplicative topological phases, demonstrating that multiplicative topological phases are an intuitive and general approach to realizing concrete models for gapless SPTs. In particular, they are naturally well-suited to realizing higher-dimensional, stable, and intrinsic gapless SPTs through combination of canonical topological insulator and semimetal models with critical gapless models in symmetry-protected tensor product constructions, opening avenues to far broader and deeper investigation of topology via short-range entanglement.
 \end{abstract}
\maketitle

Characterizing robust quantum phases of matter is at the heart of condensed matter research and of fundamental relevance for designing functional quantum devices. Paradigmatic examples without classical analogue are topological states of matter, including long-range entangled topologically ordered states\cite{levin2009, swingle2011, maciejko2012, neupert2011, santos2011, levin2011_exact, levin2012} and short-range entangled symmetry protected topological (SPT) phases. Being defined by anomalies encoding a fractionalized representation of symmetry operations\cite{chen2013, gu2014, vishwanath2013_3d, wang2013_bosonTI, burnell2014, kapustin2014symmetry, fidkowski2013, wang2014}, the latter feature gapless edge states which have been traditionally considered in gapped band structures of non-interacting fermionic topological insulators and superconductors \cite{3DTI-exp2, kane2005z2, kane2005quantum, Fu-Kane-Mele, ten-fold-way, Moorehomotopy2007} as well as strongly interacting gapped bosonic SPTs, the historically first example being Haldane's antiferromagnetic spin-1 chain \cite{haldane1983, affleck1988, chen2011_gap1Dspin, senthil2015}. In several instances, particularly in one spatial dimension, bosonic SPTs and fermionic topological band insulators are related by a non-local Jordan-Wigner mapping \cite{2D-JW-spin4,Nathanan-2020,Chen-PRB-2019,Chen-PRR-2020}.

\begin{figure}
    \centering
     \includegraphics[width=1 \columnwidth]{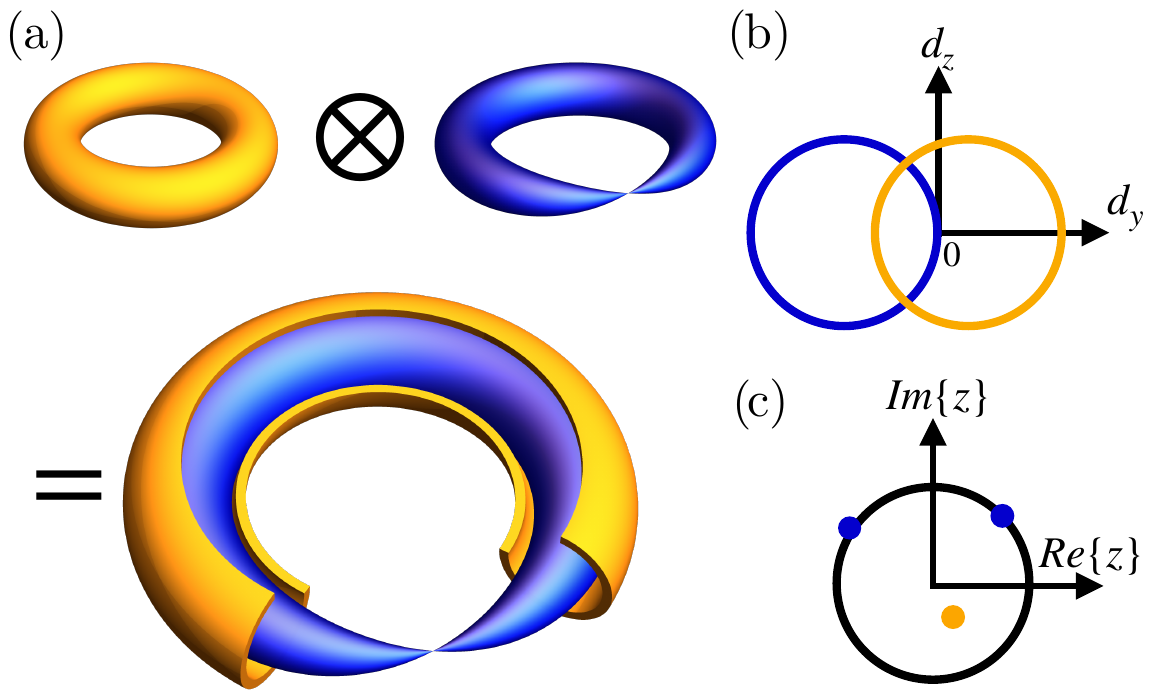}
    \caption{a) Schematic of the topological critical phase viewed as a child constructed from a topologically gapped parent (top left orange) and a critical parent (top right blue). b) Winding of the topological parents showing the critical and topological structure. c) Zeros and poles of the child topologically critical Hamiltonian in the complex plane as described in \cite{Ruben-TopoCritQuant}. }
    \label{Fig1}
\end{figure}

Recently, more exotic topological states were discovered. On the one hand, it was demonstrated that gapless SPTs (both bosonic and fermionic) exist, i.e. states of matter without a bulk gap, but which nonetheless feature a revised notion of symmetry fractionalization and (generically power-law localized) robust edge states\cite{Frank-Ruben-1d, Ruben-TopoCritQuant, scaffidi2017,Thorngren-PRB-2021}. On the other hand, multiplicative topological phases (MTPs) were revealed~\cite{cook2022mult, pal2023_semimet, pal2023_maj}. These include non-interacting fermionic phases in which the matrix Bloch Hamiltonian has symmetry-protected tensor product structure, or $h(\vec k) = h_1(\vec k) \otimes h_2(\vec k)$ is the direct product ``child'' of two ``parent'' Bloch Hamiltonians $h_{1,2}(\vec k)$ \cite{cook2022mult}, see Fig.~\ref{Fig1} a). Importantly, the ``child'' $h(\vec k)$ may display emergent topological features and describe a system of augmented spatial dimension as compared to the parents $h_{1,2}(\vec k)$ [for example the parents may describe one-dimensional systems, but the child is two-dimensional $h(k_x,k_y) = h_1(k_x) \otimes h_2(k_y)$]~\cite{pal2023_maj}.

In this work, we demonstrate that the multiplicative topological phases are an elegant and intuitive approach to concrete constructions for a myriad of gapless SPTs, including those which are higher-dimensional, stable, and intrinsic, even when the multiplicative topological phase is realized in a system without translational symmetry. While, past work on MTPs considers specifically multiplicative topological insulators~\cite{cook2022mult, pal2023_maj, pal2023_semimet}, here we complement these results by considering MTPs instead with gapless bulk. We specifically investigate the potential of multiplicative topological phases for construction of concrete Hamiltonians for gapless SPTs by considering multiplicative topological phases characterized by Bloch Hamiltonians that are symmetry-protected tensor products of two parent Hamiltonians. We choose one parent to be topologically non-trivial and gapped in the bulk, while the other parent is gapless in the bulk, as shown schematically in Fig.~\ref{Fig1} (a) and (b). We find this approach generically yields concrete constructions of gapless SPTs in the child, as illustrated in Fig.~\ref{Fig1} (c). 

To demonstrate our method, we consider three different multiplicative topological phases, $(1)$ a gapless parallel multiplicative Kitaev chain ($||$MKC), $(2)$ a gapless perpendicular multiplicative Kitaev chain ($\perp$MKC), and $(3)$ a novel \textit{Kitaev-Weyl} multiplicative topological phase, with one parent being a Kitaev chain (KC) and the other parent being a Weyl semimetal. $(1)$ provides proof of concept, $(2)$ demonstrates the potential for multiplicative topological phases to address the lack of concrete models for higher-dimensional gapless SPTs~\cite{tantivasadakarn2023}, and $(3)$ illustrates the potential of multiplicative topological phases to specifically realize stable and intrinsic gapless SPTs~\cite{thorngren_intrinsic}.

\textit{Gapless parallel multiplicative KC}---We first consider multiplicative topological phases constructed from the canonical KC Hamiltonian, reviewed in \cite{SuppMat}. We consider a Hamiltonian for a one-dimensional system $ \hat{H}=\dfrac{1}{2}\sum_k \Psi^\dagger_k h(k) \Psi_k$ with particle-hole symmetric basis $\Psi_k=\left(
c^{}_{k \uparrow},c^{}_{k \downarrow}, c^\dagger_{-k \uparrow}, c^\dagger_{-k \downarrow}
\right)^{\top}$, where $A$ and $B$ label an additional pseudospin degree of freedom (dof). We  construct the Bloch Hamiltonian $h(k)$ from two parent KC Hamiltonians as:
\begin{align}
   \quad h(k)&=h_\mu(k)\otimes h_{\mu = 1}(k)
\end{align}
where the constant $\mu$ is the chemical potential of the first parent Hamiltonian and we defined the Kitaev Hamiltonian as $h_\mu(k)=(\cos(k)+\mu)\sigma_z+\sin(k) \sigma_y$. By tuning $\mu$, the first parent Hamiltonian goes from a gapped trivial phase to a gapped topological phase if $\abs{\mu}<1$. The second parent KC chemical potential is already fixed to $1$, corresponding to gapless bulk spectrum for this parent, and corresponding gapless bulk for the resultant child Hamiltonian.

 \begin{figure}
    \centering\includegraphics[width= \columnwidth]{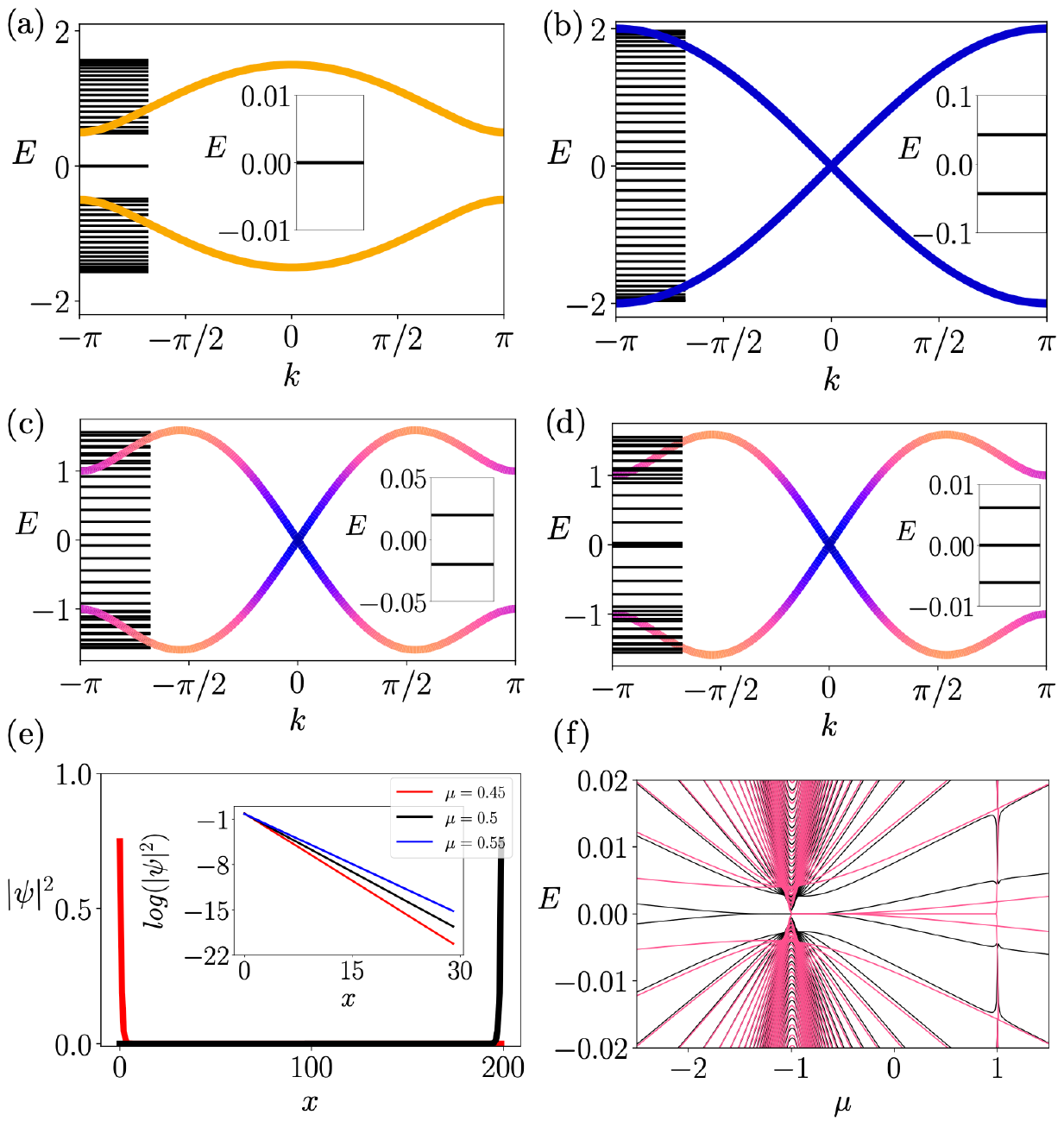}
    \caption{a) Energy spectrum of the topological Kitaev parent with periodic (orange) and open (black) boundary conditions, inset shows topological zero modes. b) Critical parent with periodic (blue) and open (black) boundary conditions with no zero modes (inset).c) $\gamma$ sector of the child Hamiltonian showing critical but no topological behaviour. d) Topological gapless $\pi$ sector of the child Hamiltonian. e) Zero modes of the child Hamiltonian localized at the boundary, inset shows exponential loclization. f) Phase diagram of the child for arbitrary chemical potential, black is the $\gamma$ sector while  pink is the $\pi$ sector, zero energy modes are only present for the region where the parent Hamiltonian is topologically non-trivial.}
    \label{Fig2}
\end{figure}
Particle-hole symmetry permits reformulation of the Hamiltonian in a Majorana basis by taking
 $c_{j,\sigma}=\frac{1}{2}(\gamma'_{j\sigma}+i\tilde{\gamma}'_{j\sigma})$, where $j$ denotes a site in real space. As the Hamiltonian also possesses a unitary symmetry corresponding to $[H(k),A]=0$ with $A=\sigma_x\tau_x$, we again change basis by defining a second set of Majorana operators $\gamma''_{j}=U\gamma'_{j}$, where $U=\frac{1}{\sqrt{2}}(\sigma_0-i\sigma_y)$, to block-diagonalize $h(k)$ into a $\gamma$ block and $\pi$ block, with the Majorana operators for the these blocks taken to be $\tilde{\gamma}_{j}=\gamma''_{j\uparrow},\gamma_{j}=\gamma''_{j\downarrow}$ and $\tilde{\pi}_{j}=\gamma''_{j\downarrow},\pi_{j}=\gamma''_{j\uparrow} $, respectively. The final second quantized Hamiltonian in this basis takes the following form $\hat{H}=\hat{H}_\gamma+\hat{H}_\pi$, where
 \begin{align}
     & \hat{H}_\pi= \dfrac{i}{2}\sum_{j}\tilde{\pi}_j\pi_{j+2}+(\mu-1) \tilde{\pi}_j\pi_{j+1}-\mu \tilde{\pi}_j\pi_j ,\label{parMKCpi}\\
     & \hat{H}_\gamma= \dfrac{i}{2}\sum_{j} (1-\mu) \tilde{\gamma}_j \gamma_j -\tilde{\gamma}_j \gamma_{j+1}+\mu \tilde{\gamma}_{j+1} \gamma_j   .
     \label{parMKCgamma}
 \end{align}
 

 We now characterize the topology of the gapless $||$MKC, 
 in which parent $1$ is a topological insulator and parent $2$ is gapless in the bulk, as shown for a representative parameter set in Fig.~\ref{Fig2} (a) and (b), respectively. Eq.~\eqref{parMKCpi} and Eq.~\eqref{parMKCgamma} are then each gapless in the bulk as shown in Fig.~\ref{Fig2} (c) and (d) 
and correspond to
 Hamiltonians previously studied by Verresen~\emph{et al.} \cite{Ruben-TopoCritQuant} in the context of gapless SPTs. In particular, 
 they showed that each of Eq.~\eqref{parMKCpi} and Eq.~\eqref{parMKCgamma} can be bosonized via Jordan-Wigner transformation to a critical chain of spins,  Eq.~\eqref{parMKCpi} realizes a critical cluster model \cite{nonne_cluster_2013} while  Eq.~\eqref{parMKCgamma} is a critical Ising model \cite{Frank-Ruben-1d}, which lies on the boundary between a ferromagnet and the AKLT phase. 
 They also define a topological invariant for one-dimensional, gapless SPTs as the number of zeros minus the order of the pole, $\nu=N_z-N_p$ inside the unit circle for the complex function $f(z)$. Here, 
 \begin{equation}
     f(z)=\sum_{\alpha} t_{\alpha} \  z^\alpha, 
     \label{fz1D}
 \end{equation} 
 where the $t_\alpha$ are the hopping amplitudes for a finite set of $\alpha$ values, and the function can be considered the unique analytical continuation of the Fourier transformed amplitudes. 
 
  Calculating Eq.~\eqref{fz1D} for the $\pi$ sector Hamiltonian Eq.~\eqref{parMKCpi}, we obtain non-trivial values whenever the gapped parent is topological, indicating the MKC realizes a gapless SPT phase. In contrast, Eq.~\eqref{fz1D} is always trivial for the $\gamma$ sector Hamiltonian Eq.~\eqref{parMKCgamma}. In correspondence with a non-trivial value for Eq.~\eqref{fz1D}, we find exponentially-localized, zero-energy bound states in the topological sector, $\pi$, as shown in Fig.~\ref{Fig2} (d) and (e), respectively. By tuning the chemical potential $\mu$, we can compute a phase diagram for the MKC by diagonalizing each sector  with open boundary conditions, as shown in Fig.~\ref{Fig2} (f), we see the zero-energy modes of the topological sector corresponding to the gapless SPT phase persist over the interval in $\mu$ over which the gapped parent is topologically non-trivial.
 
 The gapless SPT phase in the multiplicative construction therefore descends from a combination of a gapped, topologically non-trivial parent and a gapless parent. Multiplicative topological phases are therefore a very intuitive path to concrete constructions of gapless SPT phases, given the large body of literature on gapped SPT phases~\cite{senthil2015} versus gapless SPT phases \cite{Frank-Ruben-1d, Ruben-TopoCritQuant, scaffidi2017,Thorngren-PRB-2021}, in permitting construction of gapless SPT phases very directly from gapped SPT phases.

\begin{figure}
    \centering
   \includegraphics[width=\columnwidth]{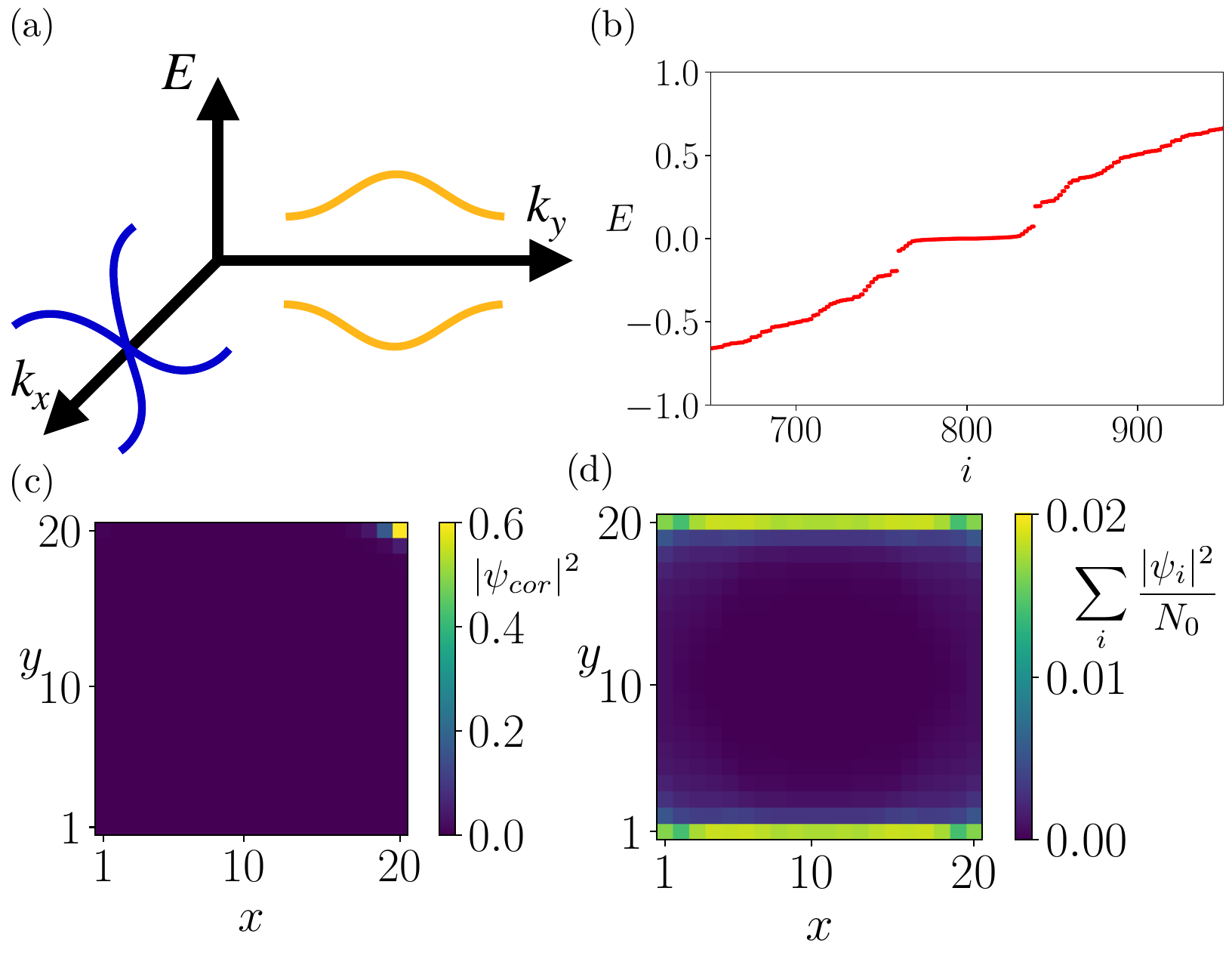}
    \caption{b) Energy spectrum as a function of eigenvalue index around zero energy in the presence of gaussian disorder with $\sigma=0.01$ averaged over 100 realizations. c) Probability density for one of the four topological corner modes present within the non-trivial region of phase space. d) Probability density for a sum of $N_0=50$ zero modes. }
    \label{Fig3}
\end{figure}

\textit{Gapless perpendicular parent KC}---The multiplicative construction allows us to easily generalize to two dimensions by simply considering two parent KCs which are perpendicular, as illustrated schematically in Fig.~\ref{Fig3} (a), to construct the perpendicular multiplicative KC~\cite{pal2023_maj}, taking our child Hamiltonian to instead be:
\begin{align}
     h(k_x,k_y)&=h_\mu(k_x)\otimes h_{1}(k_y).
     \label{perpMKC}
\end{align}

 We again exploit unitary symmetry of the Hamiltonian Eq.~\eqref{perpMKC} to block-diagonalize the Hamiltonian by going to the basis in which $A$ is diagonal (details are presented in \cite{SuppMat}). We may again change to a Majorana basis in each block of the Hamiltonian, taking our annihilation operators to be $\tilde{c}_{\textbf{k}A}=(\tilde{\pi}_\textbf{k}+i\pi_\textbf{k})/2$, and  $\tilde{c}_{\textbf{k}B}=(\tilde{\gamma}_\textbf{k}+i\gamma_\textbf{k})/2$, respectively. We may then write the Hamiltonian for one of the blocks as:
\begin{align}
      \hat{H}_\pi&= \dfrac{i}{2}\sum_{n,m}\tilde{\pi}_{n,m}\pi_{n+1,m+1}- \tilde{\pi}_{n,m}\pi_{n+1,m} \\ \nonumber
     &+\mu \tilde{\pi}_{n,m}\pi_{n,m+1}-\mu \tilde{\pi}_{n,m}\pi_{n,m} \label{perppi}.
\end{align}

The Hamiltonian for the other sector, $\hat{H}_\gamma$, is the mirror conjugate defined by the relation $\mathcal{M}_x^{-1} \pi_{n,m}\mathcal{M}_x=\gamma_{N-n,m}$, where the local action of $\mathcal{M}_x$ is 
$U_{\mathcal{M}_x} = \sigma_x$
 in the original basis. Just as in the 1D case, we can transform the Hamiltonian to a bosonized model in terms of spins. The resulting simple spin model for $\mu=0$, shown in \cite{SuppMat}, has a redefined unit cell which for open boundary conditions leads to missing edge spins reflecting the presence of edge modes for the fermionic system. The general model for arbitrary chemical potential can also be written as a spin model via the mapping described by Chun Po~\cite{2D-JW-spin4}, since the fermionic bilinears are indeed representations of $\text{Spin}(4)$. The resulting Hamiltonian is thus a three spin per site model with a local plaquette constraint, which can then be thought of as realizing a critical 2D gauge theory with topologically-protected corner modes.

For the purposes of topological characterization, we reformulate Eq.~\eqref{perppi} in the $\alpha$-chain formalism~\cite{Ruben-TopoCritQuant,Frank-Ruben-1d}, but generalizing to two dimensions. Review of this formalism and details of the two-dimensional generalization are presented in \cite{SuppMat}. The Hamiltonian $h(k_x,k_y)$ is then
\begin{align}
    \hat{H}= \sum_\alpha  \sum_\beta t_{\alpha,\beta} \ \hat{H}_{\alpha,\beta} , \quad \hat{H}_{\alpha,\beta}=\dfrac{i}{2}\sum_{n,m}\tilde{\pi}_{n,m}\pi_{n+\alpha,m+\beta}.
\end{align}

We may then generalize the topological characterization of the gapless SPTs of the $||$MKC by generalizing the previous complex function for the topological invariant, Eq.~\eqref{fz1D}, to a complex function with two arguments defined by the two dimensional Fourier transform of the Majorana tight-binding coefficients:
\begin{align}
    f(z_1,z_2)=\sum_{\alpha=-\infty}^{\infty} \sum_{\beta=-\infty}^{\infty}t_{\alpha,\beta} \  z_1^\alpha z_2^\beta. \label{fz2D}
\end{align}

In general, characterizing this function is a complicated task, since the zeros and poles are no longer strictly zero-dimensional objects, and can instead extend along any direction in the complex plane. 

A generic 2D invariant could then be calculated by considering a generalization of the 1D invariant, which takes higher-dimensional zeros and poles into consideration. Our goal is not to characterize all possible phases arising from this formalism, however, but rather to characterize gapless multiplicative topological phases. Henceforth, we therefore focus on the simplest case, described by an embedding of the 1D quantum critical topology in the 2D plane. 

We observe the multiplicative Hamiltonian \eqref{perpMKC} reduces to the 1D case previously studied if $k_x=\pm k_y$, so that we can focus on this submanifold of the Brillouin zone (BZ) if we enforce invariance of the system under mirror operations taking $x \rightarrow -x$ and $y \rightarrow -y$, respectively. The Hamiltonian \eqref{perpMKC} indeed has these symmetries. The topological characterization then employs Eq.~\eqref{fz1D} as in the 1D case, with $z_1=z_2=z$ for the $\hat{H}_{\pi}$ chain, which satisfies certain constraints on the $t_{\alpha,\beta}$, as we detail in \cite{SuppMat}. Diagonalizing the Hamiltonian for a representative parameter set, we find the low-energy spectrum is robustly gapless with high degeneracy even in the presence of disorder, as shown in Fig.~\ref{Fig3} (b). A subset of this degenerate manifold consists of zero-energy states strongly-localized at corners of the system for open boundary conditions in each of the $\hat{x}$- and $\hat{y}$-directions as shown in Fig.~\ref{Fig3} (c), hidden within the broader zero-energy manifold of states distributed on two edges of the system with these boundary conditions as shown in Fig.~\ref{Fig3} (d). 

These corner states are a consequence of the gapless SPT phase in each $\pi,\gamma$ sector. The localization of zero-energy states on the $y=1$ and $y=20$ edges derives from the extension of gapless boundary modes of the topological insulator parent defined in terms of $k_y$ along entire one-dimensional edges through combination with the $k_x$-dependent parent. The topological characterization only protects the corner modes, but the presence of the other states at zero energy and the 1D origin of the edge states reflects the nature of the phase as gapless SPT.

\begin{figure}
    \centering
   \includegraphics[width=\columnwidth]{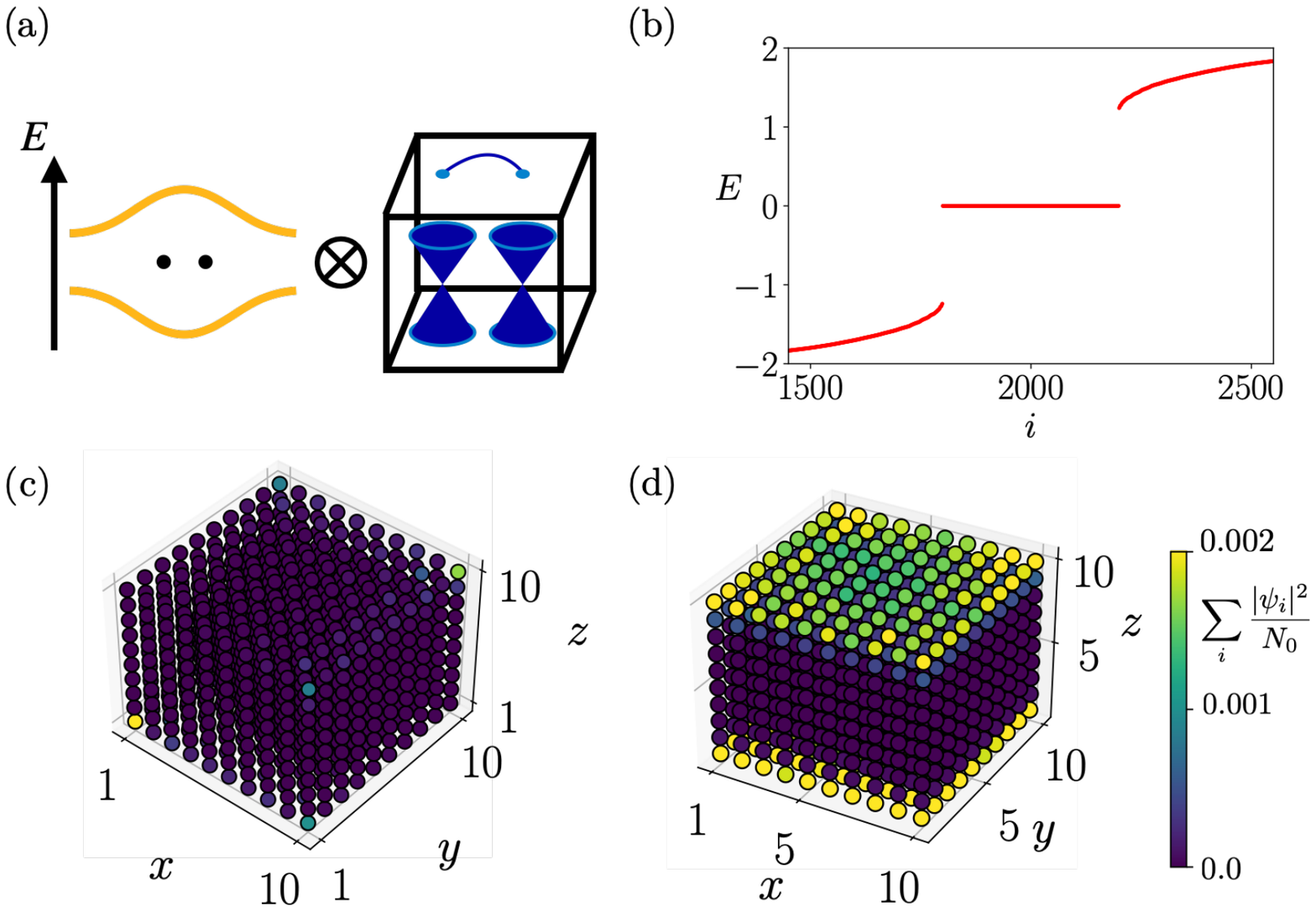}
    \caption{b) Energy spectrum as a function of eigenvalue index around zero energy in the presence of gaussian disorder with $\sigma=0.01$ averaged over 100 realizations. c) Probability density for one of the four topological corner modes present within the non-trivial region of phase space. d) Probability density for a sum of $N_0=50$ zero modes. }
    \label{Fig4}
\end{figure}

\textit{Stable and intrinsic gapless symmetry-protected topology of multiplicative topological phases}---While the multiplicative constructions realize known models for gapless SPTs in the previous two cases of the $||$MKC and $\perp$MKC, they demonstrate the key concepts required for constructing myriad novel Hamiltonians for gapless SPT phases, in particular higher-dimensional gapless SPT phases for which construction by alternate methods is highly non-trivial~\cite{tantivasadakarn2023}. Here, we illustrate the power of multiplicative topological phases in realizing more exotic gapless SPTs in higher dimensions, by combining a parent KC Hamiltonian with that of a parent Weyl semimetal (WSM) as illustrated schematically in Fig.~\ref{Fig4} (a), to realize stable and/or intrinsic gapless SPTs~\cite{thorngren_intrinsic}. This second parent topological phase is reviewed in the last section of the Supplementary Material S3. The child Hamiltonian then takes the form
\begin{align}
    h(k_x,k_y,k_z)=h_{\mathrm{WSM}}(k_x,k_y,k_z)\otimes h_{\mu}(k_z).
\end{align}

Similarly to previous sections, we consider the case of the parent KC topologically non-trivial with gapped bulk, in combination with the parent Weyl semimetal with two Weyl nodes of charge $\pm 1$ located at $\left(k_x,k_y,k_z \right)=\left(0,0,\pm \pi/2 \right)$, respectively. While stable gapless SPTs are typically realized with translational symmetry present, and intrinsic gapless SPTs are associated with strictly internal symmetry-protection in a Hamiltonian, the symmetry-protected multiplicative  construction relies on internal symmetries. We can therefore realize a child Hamiltonian without translation symmetry, yet realizing stable and intrinsic gapless SPT states, by breaking translation symmetry in the KC parent and preserving it in the WSM parent.

In this case, the previous analysis in terms of $\alpha$ chains can be extended to a three dimensional complex function $f(z_1,z_2,z_3)$ similarly to the case of the gapless $\perp$MKC. Low symmetry of the WSM parent prevent block-diagonalization of the child in this third case, however, so the invariant must also be computed with extra degrees of freedom in the unit cell, however, as considered for 1D systems in reference~\cite{Frank-Ruben-1d}.

The signatures of the non-trivial topology are easier to observe if one computes the spectrum of the previous Hamiltonian. We therefore compute the disorder-averaged spectrum for the child Hamiltonian with open-boundary conditions in each of the $\hat{x}$-, $\hat{y}$- and $\hat{z}$-directions, finding a zero-energy manifold separated from the rest of the spectrum by a finite energy gap,  Fig.~\ref{Fig4} (b).
 Similarly to the $\perp$MKC case, a subset of the zero-energy manifold consists of topological corner states of a gapless SPT phase, as shown in Fig.~\ref{Fig4} (c), while the larger zero-energy manifold consists of states localized near the $z=1$ and $z=10$ surfaces as shown in Fig.~\ref{Fig4} (d). These two-dimensional gapless surface states derive from the zero-dimensional bound states of the parent KC extended across the $xy$-plane by multiplication with states of the parent Weyl semimetal. Moreover we observe the same behaviour characteristic of gapless SPTs \cite{Frank-Ruben-1d, Ruben-TopoCritQuant, scaffidi2017,Thorngren-PRB-2021} where the system is gapless, but with open boundary conditions the bulk gap goes like $1/L$, while the edge states are exponentially close to zero as seen in Fig.\ref{Fig4} b).

\textit{Discussion \& Conclusion}---In this work, we explore the potential of multiplicative topological phases~\cite{cook2022mult, pal2023_maj, pal2023_semimet} to realize gapless symmetry-protected topological phases (SPTs) \cite{Frank-Ruben-1d, Ruben-TopoCritQuant, scaffidi2017,Thorngren-PRB-2021}. We find that they generically and intuitively realize concrete models for higher-dimensional, stable, and intrinsic gapless SPTs when the parent Hamiltonians in the tensor product of the MPT Hamiltonian consist of a topological insulator and either a critical, gapless parent or topological semimetal parent, respectively.
The variety of topologically-robust boundary modes (e.g., surface, edge, and/or corner modes) and the large symmetry of the Hamiltonians indicates that the phases studied here may serve as parent states, to be driven into a variety of topological states with more limited bulk-boundary correspondence such as higher order topological phases (e.g., gapping out 1D edge modes while preserving 0D corner modes) \cite{HOTI} or weak topological phases (e.g., preserving 2D surface modes while gapping out corner modes) \cite{Moorehomotopy2007, 10-fold-way,ten-fold-way}, which will be explored in future work. 

\textit{Acknowledgements}--- We gratefully acknowledge fruitful discussions with F.~Pollmann, T.~Scaffidi, and R.~Verressen. This research was supported in part by the National Science Foundation under Grant No. NSF PHY-1748958 and PHY-2309135, and undertaken in part at Aspen Center for Physics, which is supported by National Science Foundation grant PHY-2210452.

\bibliography{TopoQuantCritic}

\clearpage

\makeatletter
\renewcommand{\theequation}{S\arabic{equation}}
\renewcommand{\thefigure}{S\arabic{figure}}
\renewcommand{\thesection}{S\arabic{section}}
\setcounter{equation}{0}
\setcounter{section}{0}

\onecolumngrid

\newpage

\begin{center}
  \textbf{\large Supplemental material for ``Topological quantum criticality from multiplicative phases ''}\\[.2cm]
  R. Flores-Calderón,$^{1,2}$ Elio König,$^{3}$ and Ashley M. Cook$^{1,2,*}$\\[.1cm]
  {\itshape ${}^1$Max Planck Institute for Chemical Physics of Solids, Nöthnitzer Strasse 40, 01187 Dresden, Germany\\
  ${}^2$Max Planck Institute for the Physics of Complex Systems, Nöthnitzer Strasse 38, 01187 Dresden, Germany\\
  ${}^3$Max Planck Institute for Solid State Physics, Heisenberg Strasse 1, 70569 Stuttgart, Germany\\}
  ${}^*$Electronic address: cooka@pks.mpg.de\\
(Dated: \today)\\[1cm]
\end{center}

\section{One dimensional symmetry-protected multiplicative construction}

We consider the possible phases of a one dimensional chain constructed out of a tunable Kitaev chain and a critical Kitaev chain at the transition between trivial to topoological phases. The multiplicative construction implies a product of Bloch Hamiltonians which result in:
\begin{align}
    \hat{H}=\dfrac{1}{2}\sum_k \Psi^\dagger_k H(k) \Psi_k,\quad H(k)=H_\mu(k)\otimes H_c(k)= ((\cos(k)-\mu)\sigma_z+\sin(k) \sigma_y)\otimes ((\cos(k)-1)\tau_z+\sin(k) \tau_y), \label{2Dmodel}
\end{align}
where the constant $\mu$ is the chemical potential of the first parent Hamiltonian. By tuning $\mu$, the first parent Hamiltonian goes from a gapped trivial phase to a gapped topological phase if $\abs{\mu}<1$. We can then think of a basis where we have a particle-hole degree of freedom (dof) and a two-fold (pseudo)spin dof. In this case, we can transform from the complex fermion basis

\begin{align}
    \Psi_k=\begin{pmatrix}
    c^{}_{k \uparrow} && c^{}_{k \downarrow} && c^\dagger_{-k \uparrow} && c^\dagger_{-k \downarrow}
    \end{pmatrix}^T
    \label{free-basis}
\end{align}
to the Majorana fermion basis, so that when we open boundary conditions of the resulting Hamiltonian with a change of basis 
 $c_{j,\sigma}=\frac{1}{2}(\gamma'_{j\sigma}+i\tilde{\gamma}'_{j\sigma})$, we can use the chiral symmetry of the Hamiltonian to rotate by $U=\frac{1}{\sqrt{2}}(\sigma_0-i\sigma_y)$ and redefine $\gamma''_{j}=U\gamma'_{j}$. In this case, further simplification can be made by relabelling $\tilde{\gamma}_{j}=\gamma''_{j\uparrow},\gamma_{j}=\gamma''_{j\downarrow},\tilde{\pi}_{j}=\gamma''_{j\downarrow},\pi_{j}=\gamma''_{j\uparrow} $ so that the final second-quantized Hamiltonian has the form of two independent chains in this basis:
 \begin{align}
     &\hat{H}=\hat{H}_\gamma+\hat{H}_\pi,\\
     & \hat{H}_\pi= \dfrac{i}{2}\sum_{j}\tilde{\pi}_j\pi_{j+2}+(\mu-1) \tilde{\pi}_j\pi_{j+1}-\mu \tilde{\pi}_j\pi_j,\\
     & \hat{H}_\gamma= \dfrac{i}{2}\sum_{j} (1-\mu) \tilde{\gamma}_j \gamma_j -\tilde{\gamma}_j \gamma_{j+1}+\mu \tilde{\gamma}_{j+1} \gamma_j.   
 \end{align}
 We can now clearly identify from each of the two Hamiltonians the $\alpha$ chain structure~\cite{Frank-Ruben-1d} and identify the analytic continuation of the Fourier-transformed coefficients of the Majorana tight-binding Hamiltonian. The complex function we ought to characterize is then given by:
\begin{align}
    f(z)=\sum_{\alpha=-\infty}^{\infty}t_\alpha z^\alpha ,
\end{align}
where there are a finite number of $t_\alpha$ parameters linked to the $\alpha$-chains:
\begin{align}
    \hat{H}= \sum_\alpha t_\alpha \hat{H}_\alpha , \quad \hat{H}_\alpha=\dfrac{i}{2}\sum_j \tilde{\pi}_j\pi_{j+\alpha}.
\end{align}
 In our case, for the $\pi$ chain, we have $t_1=\mu-1$,  $t_2=1$, $t_0=-\mu$, while, for the $\gamma$ Majoranas, we have $t_0=1-\mu$, $t_1=-1$,  $t_{-1}=\mu$. 
 
 We now can compute the topological invariant for the $\pi$ Majoranas by calculating the number of zeros and poles as a function of the chemical potential $\mu$. A straightforward analysis using the quadratic structure of the function leads to a topological invariant $\nu=N_z-N_p$ defined as the number of zeros minus the number of poles inside the unit circle to be one or zero:
 \begin{align}
     \nu= 1,\  \abs{\mu}<1 \quad  \text{or}\quad  \nu =0, \  \text{otherwise}, 
\end{align}
while the $\gamma$ Majoranas always have a pole and a zero and so are trivial. We see now that the $\pi$ Majoranas inherit the parent Hamiltonian $H_\mu$ topological region! The analytical argument coincides with the numerical simulations, as shown in Fig. 2 f), where a zero energy mode is present exactly in the $ \abs{\mu}<1$ region. 

It is instructive to analyze the previous construction from the perspective of spins, which can be done easily by implementing a Jordan-Wigner transformation \cite{Ruben-TopoCritQuant,2D-JW-spin4}. In this case, the $\pi$ chain maps to a spin Hamiltonian, with a cluster term coming from the next-nearest neighbour term, an Ising term, and a transverse field term:
\begin{align}
    \hat{H}= \sum_j J_1 \ X_j Z_{j+1} X_{j+2}+ J_2 \ X_j X_{j+1}+J_3 \  Z_j ,
\end{align}
In our case, the values of $J_i$ which map from the original fermionic problem to the spin model are $J_3=\mu/2,J_2=(1-\mu)/2, J_1= -1/2$, the critical properties of the original fermion model map then to the critical line between the Haldane phase and the trivial paramagnet~\cite{Frank-Ruben-1d}.

\section{Two-dimensional symmetry-protected multiplicative construction}

\subsection{Model 1}
The symmetry-protected multiplicative construction~\cite{cook2022mult} allows us to easily generalize the model to two dimensions or higher by simply considering first two Kitaev chains defined in terms of two independent Cartesian coordinates, defined along perpendicular axes:
\begin{align}
     H(k_x,k_y)=H_\mu(k_x)\otimes H_c(k_y)= ((\cos(k_x)+\mu)\sigma_z+\sin(k_x) \sigma_y)\otimes ((\cos(k_y)-1)\tau_z+\sin(k_y) \tau_y) .
\end{align}
It is worth noting that, in the present form, one has again a unitary symmetry as well as spinless time reversal symmetry under $\mathcal{T}$, particle-hole symmetry under $\mathcal{P}$, mirror symmetry under the operation taking $x \rightarrow -x$ $\mathcal{M}_x$, and mirror symmetry under the operation taking $y \rightarrow -y$ $\mathcal{M}_y$. To simplify this model, let us use the fact that we have a unitary symmetry since $[H(k_x,k_y),A]=0$, with $A=\sigma_x\tau_x$. This means we can take the eigenvectors of this operator forming a unitary matrix $U_A$ to block diagonalize the Hamiltonian into
\begin{align}
     \hat{H}=\dfrac{1}{2}\sum_\textbf{k} \Psi^\dagger_\textbf{k} H(k_x,k_y) \Psi_\textbf{k}=\dfrac{1}{2}\sum_\textbf{k} \tilde{\Psi}^\dagger_\textbf{k}U_A H(k_x,k_y) U_A^\dagger \tilde{\Psi}_\textbf{k} ,
\end{align}
where we have the original basis $\Psi_\textbf{k}=\left(
c_{\textbf{k}A}, c_{\textbf{k}B}, c^\dagger_{-\textbf{k}A} ,c^\dagger_{-\textbf{k}B}
\right)^T$ and redefined $\tilde{\Psi}_\textbf{k}=U_A \Psi_\textbf{k}$ let us go now to a Majorana basis by using the definitions of $\tilde{c}_{\textbf{k}A}=(\tilde{\pi}_\textbf{k}+i\pi_\textbf{k})/2$, and  $\tilde{c}_{\textbf{k}B}=(\tilde{\gamma}_\textbf{k}+i\gamma_\textbf{k})/2$. This can be implemented with a new matrix $U_{\text{maj}}$ we obtain then two decoupled models of Majoranas:
\begin{align}
 \hat{H}_\pi=\sum_\textbf{k}\dfrac{1}{4}\begin{pmatrix}
  \tilde{\pi}_{-\textbf{k}} && \pi_{-\textbf{k}}
 \end{pmatrix}
\begin{pmatrix}
 0 & i\left(-1+e^{i k_y}\right) (e^{ i k_x}+\mu ) \\
 -i\left(-1+e^{-i k_y}\right) (e^{ -i k_x}+\mu ) & 0 \\
\end{pmatrix}\begin{pmatrix}
  \tilde{\pi}_\textbf{k} \\ \pi_{\textbf{k}}
 \end{pmatrix},\\
 \hat{H}_\gamma=\sum_\textbf{k}\dfrac{1}{4}\begin{pmatrix}
  \tilde{\gamma}_{-\textbf{k}} && \gamma_{-\textbf{k}}
 \end{pmatrix}
\begin{pmatrix}
 0 & i\left(-1+e^{i k_y}\right) (e^{ -i k_x}+\mu ) \\
-i\left(-1+e^{-i k_y}\right) (e^{ i k_x}+\mu ) & 0 \\
\end{pmatrix}\begin{pmatrix}
  \tilde{\gamma}_\textbf{k} \\ \gamma_{\textbf{k}}
 \end{pmatrix} .
\end{align}

Transforming to a Majorana representation just as before with the same basis of fermions, one obtains two decoupled Hamiltonians:

\begin{align}
     &\hat{H}=\hat{H}_\gamma+\hat{H}_\pi,\\
     & \hat{H}_\pi= \dfrac{i}{2}\sum_{n,m}\tilde{\pi}_{n,m}\pi_{n+1,m+1}- \tilde{\pi}_{n,m}\pi_{n+1,m}+\mu \tilde{\pi}_{n,m}\pi_{n,m+1}-\mu \tilde{\pi}_{n,m}\pi_{n,m},\\
     & \hat{H}_\gamma= \dfrac{i}{2}\sum_{n,m}\tilde{\gamma}_{n,m}\gamma_{n-1,m+1}- \tilde{\gamma}_{n,m}\gamma_{n-1,m}+\mu \tilde{\gamma}_{n,m}\gamma_{n,m+1} -\mu \tilde{\gamma}_{n,m}\gamma_{n,m} .
\end{align}

For the two-dimensional model, one can think of a generalization of the previous complex function to a complex function with two arguments, defined from the two-dimensional Fourier transform of the Majorana tight-binding coefficients:

\begin{align}
    f(z_1,z_2)=\sum_{\alpha=-\infty}^{\infty} \sum_{\beta=-\infty}^{\infty}t_{\alpha,\beta} \  z_1^\alpha z_2^\beta  ,\label{fz2d}
\end{align}
which can be defined using a generalization of the $\alpha$-chains to 2D as:
\begin{align}
    \hat{H}= \sum_\alpha  \sum_\beta t_{\alpha,\beta} \ \hat{H}_{\alpha,\beta} , \quad \hat{H}_{\alpha,\beta}=\dfrac{i}{2}\sum_{n,m}\tilde{\pi}_{n,m}\pi_{n+\alpha,m+\beta} .
\end{align}

Characterizing the complex function is now much more complicated, since for more than one complex variable, the function's zeros and poles are not generally zero-dimensional subsets. In this case, the analysis can be simplified if one uses the mirror symmetries. The chains are actually related by the mirror operation $\mathcal{M}_x$, such that $\mathcal{M}_x^{-1} \gamma_{n,m}\mathcal{M}_x=\pi_{N-n,m}$. The mirror symmetries of the system imply $k_x\rightarrow-k_x$ and $k_y\rightarrow -k_y$ so that the previously-defined function in equation \eqref{fz2d} satisfies $z_1=e^{ik_x}\rightarrow e^{-ik_x}=\dfrac{1}{z_1}$ similarly,$z_2=e^{ik_y}\rightarrow e^{-ik_y}=\dfrac{1}{z_2}$, which means that the $\mathcal{M}_x,\mathcal{M}_y$ symmetries imply $f(z_1,z_2)=f(1/z_1,z_2)$. Or, equivalently:

\begin{figure}
    \centering
   \includegraphics[width=0.3\textwidth]{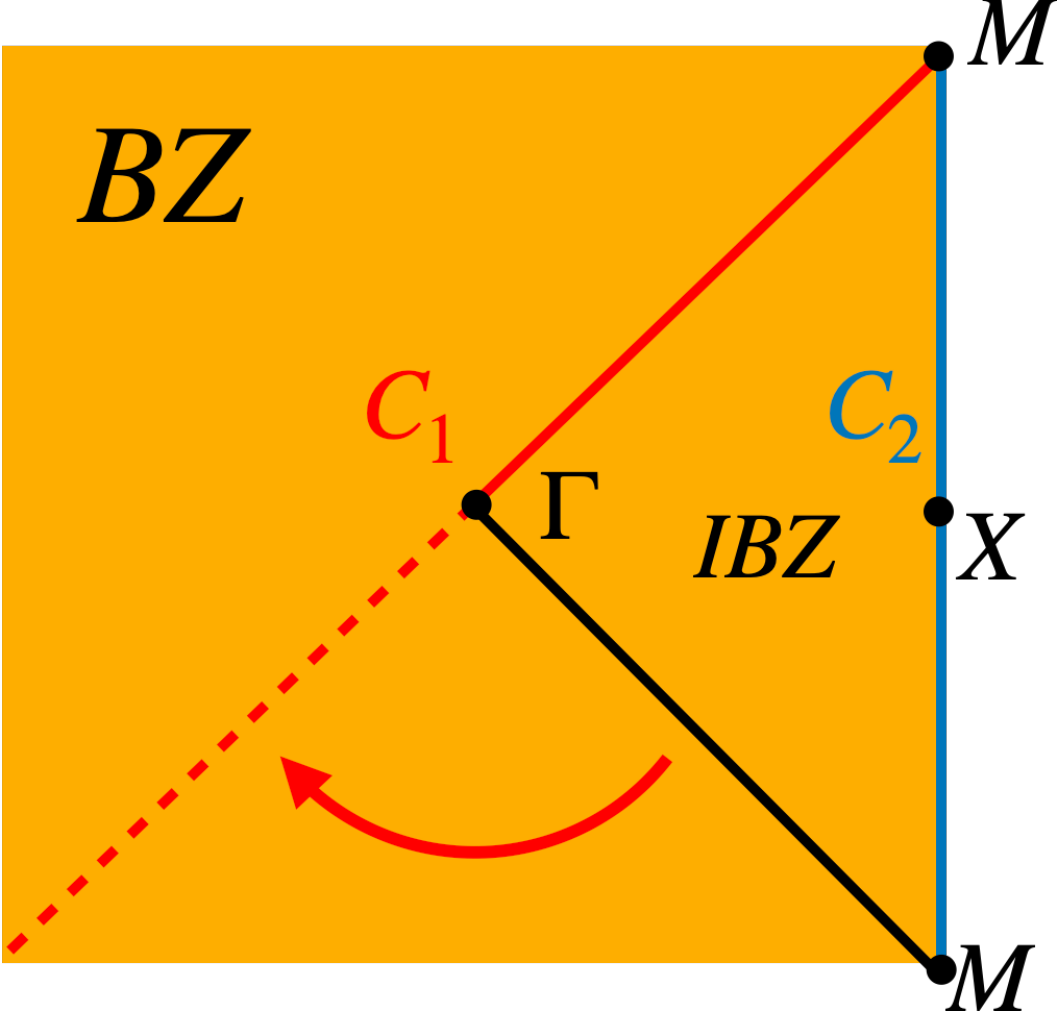}
    \caption{Schematic of the Brillouin Zone (BZ) and Irreducible BZ (IBZ) enforced by $C_4T$ used for the characterization of the Majorana edge modes of the 2D model.}
    \label{FigIBZ}
\end{figure}

\begin{align}
   \sum_{\alpha=-\infty}^{\infty} \sum_{\beta=-\infty}^{\infty}t_{\alpha,\beta} \  z_1^\alpha z_2^\beta =\sum_{\alpha=-\infty}^{\infty} \sum_{\beta=-\infty}^{\infty}t_{\alpha,\beta} \  z_1^{-\alpha} z_2^\beta=\sum_{\alpha=-\infty}^{\infty} \sum_{\beta=-\infty}^{\infty}t_{-\alpha,\beta} \  z_1^{\alpha} z_2^\beta ,
\end{align}
which means then that we have a constraint on the possible values of the hopping parameters, meaning: $t_{\alpha,\beta}=t_{-\alpha,\beta}$ as well as $t_{\alpha,\beta}=t_{\alpha,-\beta}$ and so $t_{\alpha,\beta}=t_{-\alpha,-\beta}$. By virtue of this relations we may for simplicity focus on the case of positive $\alpha,\beta$.  We note also a $C_4T$ symmetry of each Hamiltonian which we use to define the irreducible Brillouin Zone (IBZ). The boundary of the IBZ can be taken to be the right triangle of the BZ as shown in Fig \ref{FigIBZ}. We define relevant high-symmetry points in the IBZ of $\Gamma = \left( k_x=0,k_y=0\right)$ and $M = \left( k_x=\pi,k_y=\pi\right)$, and consider two curves along the boundary $\partial (IBZ)=\mathcal{C}_1 \cup \mathcal{C}_2 $, where $\mathcal{C}_2$ goes from $M$ to $X$ and then $M$ again,  parameterized by $k_x=\pi$, $k_y=t$ and $\mathcal{C}_1$ covers the diagonal $\Gamma$ to $M$ which is parametrized by $k_x=k_y=t$ where $t\in (-\pi,\pi)$. We focus on the full diagonal path $C_1$ because thanks to $C_4T$ the antidiagonal part of $\partial(IBZ)$ in black , Fig.\ref{FigIBZ} , can be mapped to the dashed red line.  We can define the equivalent of the one dimensional function if we take $e^{ik_x}=-1$ and $e^{it}\rightarrow z$ for $\mathcal{C}_1$ and $e^{ik_x}=e^{ik_y}=\rightarrow z$ for $\mathcal{C}_2$, which leads to generically defining:
\begin{align}
    f_1(z)=\sum_{\alpha,\beta} t_{\alpha,\beta} (-1)^\beta z^\alpha, \qquad f_2(z)=\sum_{\alpha,\beta} t_{\alpha,\beta} (-1)^\alpha z^\beta, \qquad  f_+(z)=\sum_{\alpha,\beta} t_{\alpha,\beta} z^{\alpha+\beta}. 
\end{align}

We conjecture that under this assumptions the topological invariant is again  $\nu_+=N^+_z-N^+_p$ which counts the number of zeros in $f_+$ minus the order of the pole in $f_+$. Similarly, we also define $\nu_1=N^1_z-N^1_p$, and $\nu_2=N^2_z-N^2_p$. Let us now prove the topological invariant $\nu_+$ indeed counts the number of Majorana modes on the edge. Following the reasoning in the one dimensional case, we define a localized Majorana mode if:
\begin{align}
    \pi_1^{(i)}=\sum_{l,k\geq 1} b^{(i)}_{l,k} \pi_{l,k},\quad  [\pi_1^{(i)},\hat{H}]=0,\quad  \{\pi_1^{(i)},\pi_1^{(j)} \}=2\delta_{i,j}, \quad  \abs{b^{(i)}_{l,k}} \sim \abs{z_i^1}^l \abs{z_i^2}^k 
\end{align}
Let us denote by $z_i$ the $\nu$ largest zeros of $f_+$. We also first assume that $N_p^i=0$ which implies neither $\alpha$ nor $\beta$ take negative values in the $t_{\alpha,\beta}$ hoppings. The zero-energy condition for localized Majorana modes is implemented by considering the commutator with the full Hamiltonian:
\begin{align}
    [\pi_1^{(i)},\hat{H}]&= \dfrac{i}{2}\sum_{n,m} \sum_{\alpha,\beta} t_{\alpha,\beta}\sum_{l,k\geq 1} b^{(i)}_{l,k} \  [\pi_{l,k},\tilde{\pi}_{n,m}\pi_{n+\alpha,m+\beta}] =  \dfrac{i}{2}\sum_{n,m} \sum_{\alpha,\beta} t_{\alpha,\beta}\sum_{l,k\geq 1} b^{(i)}_{l,k} \  (-2 \delta_{l,n+\alpha}\delta_{k,m+\beta}\tilde{\pi}_{n,m} \pi_{l,k} \pi_{n+\alpha,m+\beta})\\
    =& -i \sum_{n,m} \sum_{l,k} t_{l-n,k-m} b^{(i)}_{l,k} \tilde{\pi}_{n,m} =-i \sum_{n,m} C_{n,m} \tilde{\pi}_{n,m}, \quad C_{n,m}= \sum_{\alpha,\beta\geq 1}  b^{(i)}_{\alpha,\beta} t_{\alpha-n,\beta-m} .
\end{align}
We need $C_{n,m}$ to vanish for each $n$, $m$ for the Majorana mode to be a zero mode. We notice all coefficients of the complex function are contained in $C_{n,m}$, since $t_{\alpha,\beta< 0}=0$, assuming that there are no poles. We choose  $b^{(i)}_{l,k}= z_i^{l+k-2} $, which means $C_{n,m}=z_i^{n+m-2} f_+(z_i) =0$. If the complex number is real, this will define a real Majorana that is normalizable, since  we can calculate $ \{\pi_1^{(i)},\pi_1^{(i)} \}= \sum_{n,m\geq 1}^{\infty}(b^{(i)}_{n,m})^2=2 \sum_{n=1}^{\infty} \sum_{m=1}^{\infty} (z_i^2)^{n-1}  (z_i^2)^{m-1}  = \dfrac{2}{1-z_i^2}\dfrac{1}{1-z_i^2}\neq 0$.
In order to obtain Majorana zero modes we need $\pi_1^{(i)}$ to be Hermitian, $\pi_1^{(i)}$ will also commute with time reversal $T^2=1$ if the coefficients $b^{(i)}_{l,k}$ are real. Because of the hermiticity of the Majorana operators, we notice that, even if $z_i$ is complex, we can build a real weight by using $b^{(i)}_{l,k}=z_i^{l+k-2}\pm \bar{z}_i^{l+k-2}$, which will give two solutions based on the two distinct zeros $z_i$ and $\bar{z}_i$.\\

Let's consider the case now of only poles in the complex functions i.e. $\nu_+,\nu_1,\nu_2<0$ and $N_z^{j}=0$. In this case, we can map it to the previous case, in which there are no poles, by noting that, in this case, all coefficients of $\alpha,\beta$ are negative ($t_{\alpha,\beta \geq 0} =0$) and so redefining:
\begin{align}
    \hat{H}&= \sum_{\alpha=-\infty}^{\infty}  \sum_{\beta=-\infty}^{\infty}  t_{\alpha,\beta} \ \hat{H}_{\alpha,\beta}=\sum_{\alpha=1}^{\infty}  \sum_{\beta=1}^{\infty}  t_{-\alpha,-\beta} \ \hat{H}_{\alpha,\beta} , \quad \hat{H}_{\alpha,\beta}=\dfrac{i}{2}\sum_{n,m}\tilde{\pi}_{n,m}\pi_{n-\alpha,m-\beta}\\
     f_1'(z)&=\sum_{\alpha,\beta\geq 1} t_{-\alpha,-\beta} (-1)^\beta z^\alpha, \qquad f_2'(z)=\sum_{\alpha,\beta\geq 1} t_{-\alpha,-\beta} (-1)^\alpha z^\beta, \qquad  f_+'(z)=\sum_{\alpha,\beta \geq 1} t_{-\alpha,-\beta} z^{\alpha+\beta}, 
\end{align}
The correct zero-energy condition requires now fulfillment of $C_{n,m}= \sum_{\alpha,\beta\geq 1}  b^{(i)}_{\alpha,\beta} t_{n-\alpha,m-\beta}$. Choosing the largest zeros $z_i$ of $f_+'$, we get that $b^{(i)}_{\alpha,\beta}=z^{\alpha+\beta-2}$ is again a solution since $C_{n,m}=z_i^{n+m-2}\sum_{l\geq 1-m,k \geq 1-m}t_{-l,-k}z_i^{l+k}=z_i^{n+m-2}f_+'(z_i)=0$. The second-to-last equality derives from the previous condition on the hopping parameters. The state is also normalizable since it has the same functional form as before. We note that, for the function we consider, the zeros from the function with an inverted argument are actually related to our previous function, since, with the condition $ t_{\alpha,\beta \geq 0} =0$, we have $f_+'(1/z)=\sum_{\alpha,\beta \geq 1} t_{-\alpha,-\beta} z^{-\alpha-\beta}=\sum_{\alpha,\beta } t_{\alpha,\beta} z^{\alpha+\beta}=f_+(z)$. This means the invariant, in this case, is the number of zeros outside the unit circle of $f_+(1/z)$ . Now, since we can always write $f_+(z)=1/z^{N_p^+}\prod_i (z-z_i)$, it is clear that $f_+(1/z)=1/z^{N-N_p^+}\prod_i (z-1/z_i)$, with $N$ being the total number of zeros. Let us take the number of zeros on the unit circle to be $2c$, with $c$ possibly half integer. Then, since the number of zeros of $f_+(1/z)$ outside the unit circle is the same as the number of zeros inside the unit circle for $f_+(z)$, the topological invariant of this function is $\nu'=N_z'-N_p'=(N-N_z^+-2c)-(N-N_p^+)$. Therefore $\nu'=-\nu_+ -2c>0$, since we supposed there were only zeros for the function with inverted coordinates. We conclude, from this reasoning, that the system has $\abs{\nu_+ +2c}=\abs{-N_p^+ +2c}$ localized corner modes.

\subsection{Jordan-Wigner Transformation}

As the Hilbert space at each lattice site is spanned by four Majorana operators which can be used to compose bilinears $\tilde{\gamma}_j\gamma_i,\tilde{\pi}_{j\pi} $ we have a Hilbert space of $\text{Spin}(4) \cong \text{SU}(2)\times \text{SU}(2) $, additionally the ground state projector has for multiplicative Hamiltonians the structure  $\text{Spin}(4) \cong \text{SU}(2)\times \text{SU}(2)$ \cite{cook2022mult}. It is possible then to map the original fermion Hamiltonian to a spin Hamiltonian via a two-dimensional (2D) Jordan Wigner transformation developed specifically for $\text{Spin}(4)$ Hamiltonians in \cite{2D-JW-spin4}. We also present a Jordan-Wigner transformation, which works in a more conventional form via 2D strings for the case of $\mu=0$. 

\subsubsection{Spin(4) Jordan Wigner transformation}

For general $\mu$ and focusing only on the $\pi$ Majoranas, we can use the mapping presented in \cite{2D-JW-spin4} where we identify $\pi\rightarrow-\gamma^2,\ \tilde{\pi}\rightarrow \gamma^{1},\  \gamma \rightarrow -\gamma^4,\  \tilde{\gamma} \rightarrow  \gamma^3$ following their notation for labelling the local Majorana operators. The transformation presented in this reference makes use of the exceptional isomorphism mentioned before by representing the Majorana bilinears on each site $\tilde{\gamma}^i\gamma^j$, which are acted upon by $\text{Spin(4)}$, as a product of two degrees of freedom each describing one $\text{SU}(2)$: one degree of freedom is the charge sector and the other is the spin sector. The Hilbert space of both charge and spin sectors written in terms of new fermionic variables, partons, becomes of dimension $2^4$ while the original Hilbert space dimension was four. One therefore needs to further restrict the partons to a subspace dictated by the parton parity operator $\hat{\Gamma}\rightarrow -1$ reflecting the original fermionic degrees of freedom. Restricting the occupation of the fermions effectively leaves bosonic variables $\Theta$. The remaining unphysical degrees of freedom unaccounted for are used to fulfill fermion commutation relations between different lattice sites. We proceed by first mapping the Majorana bilinears to the bosonic operators. We then transform to the spin representation, which uses the parton constraint.  The easiest terms to transform first are the on-site bilinears, we have:
\begin{align}
    i\tilde{\pi}_{n,m}\pi_{n,m}\rightarrow-i\gamma^{1}_{n,m}\gamma^2_{n,m}\rightarrow -\Theta^{12}_{n,m}, \quad i\tilde{\gamma}_{n,m}\gamma_{n,m}\rightarrow-i\gamma^{3}_{n,m}\gamma^4_{n,m}\rightarrow -\Theta^{34}_{n,m},
\end{align}
 where the $\Theta$ bosonic operator comes from the alternate representation of the $SU(2)$ charge and spin sectors. Similarly, for the hoppings, we have:
 \begin{align}
      i\tilde{\pi}_{n,m}\pi_{n+1,m}\rightarrow-i\gamma^{1}_{n,m}\gamma^2_{n+1,m}\rightarrow -\Lambda^{11}_{n,m}\Lambda^{22}_{n+1,m},\quad   i\tilde{\gamma}_{n,m}\gamma_{n-1,m}\rightarrow-i\gamma^{3}_{n,m}\gamma^4_{n-1,m}\rightarrow \Lambda^{33}_{n,m}\Lambda^{44}_{n-1,m}\\
      i\tilde{\pi}_{n,m}\pi_{n,m+1}\rightarrow-i\gamma^{1}_{n,m}\gamma^2_{n,m+1}\rightarrow -\Lambda^{11}_{n,m}\Lambda^{22}_{n,m+1},\quad   i\tilde{\gamma}_{n,m}\gamma_{n,m+1}\rightarrow-i\gamma^{3}_{n,m}\gamma^4_{n,m+1}\rightarrow -\Lambda^{33}_{n,m}\Lambda^{44}_{n,m+1} ,
 \end{align}
   where $\Lambda$ is an operator designed to connect the Hilbert spaces of each site so as to not increase the degrees of freedom, this will also be mapped to spins in the end. Proceeding with the diagonal coupling which is further range we obtain:
 \begin{align}
     i\tilde{\pi}_{n,m}\pi_{n+1,m+1}\rightarrow-i\gamma^{1}_{n,m}\gamma^2_{n+1,m+1}\rightarrow- \Lambda^{14}_{n,m}\Phi^{32}_{n+1,m}\Lambda^{21}_{n+1,m+1},\\
     i\tilde{\gamma}_{n,m}\gamma_{n-1,m+1}\rightarrow-i\gamma^{3}_{n,m}\gamma^4_{n-1,m+1}\rightarrow \Lambda^{33}_{n,m}\Phi^{42}_{n-1,m}\Lambda^{41}_{n-1,m+1},
 \end{align}
 here we also used the $\Phi$ operator which can be written in terms of the other previously defined operators, $\Theta,\Lambda$, as shown in ref. \cite{2D-JW-spin4} . These operators can be mapped to a lattice of three spins as mentioned before per unit cell which we write compactly in the $X^i,Y^i,Z^i$ notation \cite{2D-JW-spin4} , where each operator is a Pauli matrix. The mapping is then:
\begin{align}
    \Theta^{12}_{n,m}\rightarrow-Z^1_{n,m}, \quad \Theta^{34}_{n,m}\rightarrow-Z^1_{n,m}Z^2_{n,m},\\
    \Phi^{42}_{n-1,m}\rightarrow - Z^2_{n-1,m} Y^3_{n-1,m}, \quad \Phi^{32}_{n+1,m}\rightarrow X^3_{n+1,m},
\end{align}
while the $\Lambda$ operators are the following:
\begin{align}
     \Lambda^{11}_{n,m} \rightarrow X^1_{n,m} X^2_{n,m} X^3_{n,m}, \quad  \Lambda^{22}_{n,m} \rightarrow Y^1_{n,m} X^2_{n,m} Y^3_{n,m}, \quad  \Lambda^{33}_{n,m}\rightarrow Z^1_{n,m}X^2_{n,m}Z^3_{n,m}, \quad  \Lambda^{44}_{n,m}\rightarrow X^2_{n,m} \\
    \Lambda^{21}_{n+1,m+1} \rightarrow Y^1_{n+1,m+1} X^2_{n+1,m+1} X^3 _{n+1,m+1}, \quad 
    \Lambda^{14}_{n,m} \rightarrow X^1_{n,m} Y^2_{n,m}, \quad \Lambda^{41}_{n-1,m+1} \rightarrow -Y^2_{n-1,m+1}X^3_{n-1,m+1}.
\end{align}
The Hamiltonian of the $\pi$ Majoranas becomes then:
\begin{align}
    H_{\pi}=  \dfrac{1}{2}\sum_{n,m}\left( - X^1_{n,m} Y^2_{n,m}X^3_{n+1,m}Y^1_{n+1,m+1} X^2_{n+1,m+1} X^3 _{n+1,m+1}+X^1_{n,m} X^2_{n,m} X^3_{n,m}Y^1_{n+1,m} X^2_{n+1,m} Y^3_{n+1,m}\right. \\
    \left.-\mu X^1_{n,m} X^2_{n,m} X^3_{n,m+1}Y^1_{n,m+1} X^2_{n,m+1} Y^3_{n,m+1}-\mu Z^1_{n,m}\right).
\end{align}
We must also impose a constraint that reduces the number of degrees of freedom of the three qubits to the physical space. This constraint takes the form of a flux attachment, since the second qubit encodes the onsite fermion parity, from Wen's plaquette model \cite{Wen-2003-PRL,2D-JW-spin4}:
\begin{align}
Y^3_{n,m}X^3_{n+1,m}Y^3_{n+1,m+1}X^3_{n,m+1} = Z^2_{n,m}Z^2_{n,m+1}    
\end{align}
Although this model works for all values of the chemical potential $\mu$, it is also a complicated gauge theory because of the above mentioned constraint.

\subsubsection{Jordan Wigner transformation with a string}

We may study the special case of $\mu=0$ in a more physically illustrative way by doing a normal Jordan-Wigner (JW) transformation, but in 2D. We must therefore define a Jordan-Wigner string, which wraps around  the $y$ axis of the plane in a consistent way,  meaning all sites are accounted for appropriately.  

\begin{figure}
    \centering
     \includegraphics[width=0.4 \columnwidth]{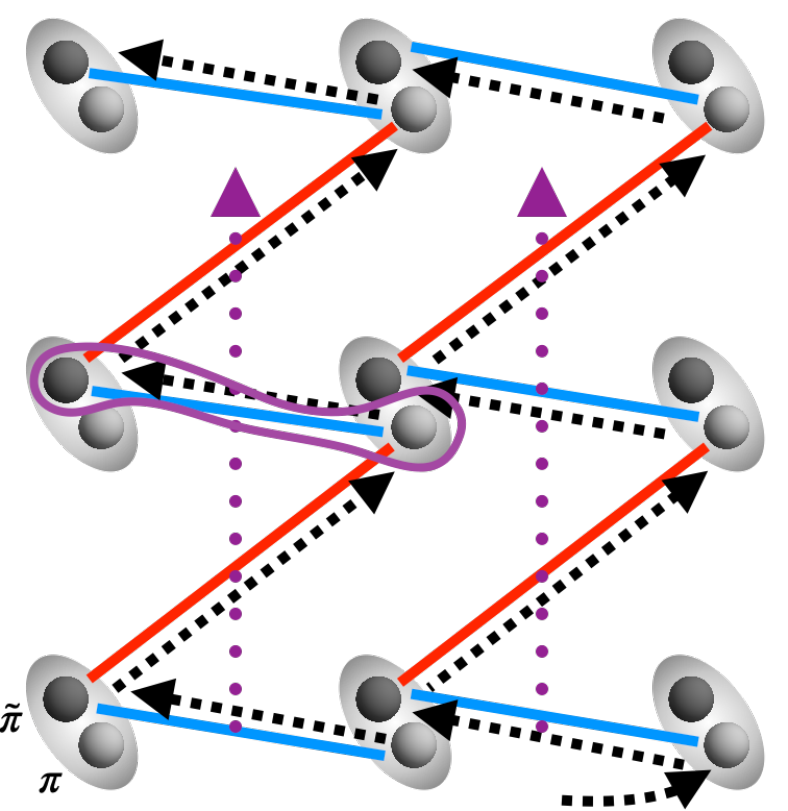}
    \caption{Schematic of $\pi$ Majoranas in the lattice and the Jordan Wigner strings, blue lines indicate a hopping of $+1$ while red ones $-1$. The gray unit cell is the original unit cell of the lattice while the purple one is a redefinition used for the Jordan Wigner string.}
    \label{FigJW2D}
\end{figure}

The lattice of $\pi$  Majoranas is shown in Fig.~\ref{FigJW2D}: we see there are two possible ways of defining the Jordan Wigner string. The easiest way is by redefining the unit cell to contain the $\tilde{\pi}_{n,m}$ Majorana together with the $\pi_{n+1,m}$ Majorana as shown in purple in Fig.~\ref{FigJW2D}. The Jordan-Wigner string is then the one going up the lattice in the $y$ direction, this is done for all $n$ which leads to an extensive 
number of decoupled Ising Hamiltonians of the form
\begin{align}
    \hat{H}= \sum_{n,m} Y_{n,m}Y_{n,m+1}-Z_{n,m},
\end{align}
where we have assumed a JW decomposition that identifies $\pi$ with a string ending in a Pauli $X$ and  $\tilde{\pi}$ with a string ending in $Y$. Although the model looks trivial in this basis, it is worth noting that, when the lattice terminates  in the $x$ direction, an extensive 
number of zero Majoranas are left at the boundary since the unit cell was modified from the gray to the purple one Fig.~\ref{FigJW2D} . This situation is analogous to that of the Su-Schrieffer-Heeger (SSH) chain at the limit of complete dimerization \cite{10-fold-way,ten-fold-way}. In the case of the SSH chain, it is also possible to redefine the unit cell, such that the SSH Hamiltonian corresponds to an infinite number of decoupled zero-dimensional systems. The lattice termination leads to edge modes at the boundary in what looks like a trivial way. In our case the 2D model \eqref{2Dmodel} we study is a generalization where the decoupled zero-dimensional systems are now decoupled 1D critical chains.
\subsection{Three-dimensional Kitaev-Weyl multiplicative topological phse}

To characterize topological quantum criticality of the Bloch Hamiltonian with symmetry-protected tensor product structure, with one parent Hamiltonian being that of a Weyl semimetal $H_{\text{WSM}}(k_x,k_y,k_z)$, and the other parent Hamiltonian being that of a Kitaev chain $H_{\mu}(k_z)$, as
\begin{align}
     &H(k_x,k_y,k_z)=H_{\text{WSM}}(k_x,k_y,k_z)\otimes H_{\mu}(k_z)
     \\&=(( \cos(k_x)+\cos(k_y)+\cos(k_z)-1)\sigma_z+\sin(k_x) \sigma_x+\sin(k_y) \sigma_y)\otimes((\cos(k_z)-\mu)\tau_z+\sin(k_z) \tau_y),
\end{align}
we first consider the natural generalization of the complex function used to characterise one-dimensional and two-dimensional gapless SPTs of the parallel and perpendicular multiplicative Kitaev chain,

\begin{align}
    f(z_1,z_2,z_3)=\sum_{\alpha=-\infty}^{\infty} \sum_{\beta=-\infty}^{\infty}\sum_{\gamma=-\infty}^{\infty} t_{\alpha,\beta,\gamma} \  z_1^\alpha z_2^\beta z_3^\gamma,
\end{align}
with the coefficients now defined through generalization of the $\alpha$ chains to 3D:
\begin{align}
    \hat{H}= \sum_\alpha  \sum_\beta \sum_\gamma t_{\alpha,\beta,\gamma} \ \hat{H}_{\alpha,\beta,\gamma} , \quad \hat{H}_{\alpha,\beta,\gamma}=\dfrac{i}{2}\sum_{n,m}\tilde{\pi}_{n,m,l}\pi_{n+\alpha,m+\beta,l+\gamma}.
\end{align}

The child Hamiltonian possesses a large symmetry group, including time-reversal $\mathcal{T}$, parity $\mathcal{P}$, and spatial inversion $\mathcal{I}$, as well as mirrors $\{\mathcal{M}_i\}$ and $i \in \{x, y, z\}$, corresponding to invariance under the operation $i \rightarrow -i$.  Similarly to the two-dimensional case, these symmetries reduce topological characterisation of the full three-dimensional system to topological characterisation of lower dimensional submanifolds of the Brillouin zone. As this Hamiltonian lacks a local unitary symmetry, it is not possible to block-diagonalise the Hamiltonian, requiring generalisation from a one site unit cell to multiple site unit cells~\cite{Frank-Ruben-1d}. Then we examine $T_{\alpha,\beta,\gamma} \in \mathbb{R}^N \times \mathbb{R}^N$ and compute the determinant $f(z)=\text{det}(\sum_{\alpha,\beta,\gamma}T_{\alpha,\beta,\gamma}z^{\alpha+\beta+\gamma})$ for $N$ sites in the unit cell. The topological invariant is again the number of zeros minus the order of the pole $\nu=N_z-N_p$: we may proceed as in characterization of one-dimensional systems for the submanifold under consideration. Given the complexity of the model despite its symmetries, we perform this characterization numerically.

\end{document}